\documentclass[11pt]{article}
\usepackage{amsmath,amssymb,amscd,latexsym,amsthm,mathrsfs}
\usepackage[unicode]{hyperref}
\usepackage{hypbmsec,longtable}
\usepackage{fancyhdr}
\ifx\pdfoutput\undefined\else
\hypersetup{
colorlinks=true,
linkcolor=blue,
citecolor=blue,
urlcolor=blue,
filecolor=blue,
bookmarksnumbered=true,
pdfstartview=FitH,
pdfhighlight=/N
}
\fi

\newtheorem{theorem}{Theorem}

\theoremstyle{remark}

\def\m{{\operatorname{m}}}

\usepackage{cite}
\def\e{{\rm e}}
\def\m{{\rm m}}
\renewcommand{\author}[1]{\large\rm #1\\ \bigskip}
\newcommand{\address}[1]{{\normalsize\it #1\\}\bigskip}
\renewcommand{\title}[1]{\bigskip\bigskip\Large\bf #1\bigskip\bigskip\\}

\begin{document}


\vglue .3 cm

\begin{center}

\title{Spanning tree generating functions and Mahler measures}
\author{              Anthony J. Guttmann\footnote[1]{email:
                {\tt tonyg@ms.unimelb.edu.au}}}
\address{ ARC Centre of Excellence for\\
Mathematics and Statistics of Complex Systems,\\
Department of Mathematics and Statistics,\\
The University of Melbourne, Victoria 3010, Australia}
\author{              Mathew D Rogers\footnote[2]{email:
                {\tt mathewrogers@gmail.com}}}
\address{
Department of Mathematics and Statistics,\\
Universit\'e de Montr\'eal, Montr\'eal, Qu\'ebec, H3C\,3J7, Canada}

\end{center}
\setcounter{footnote}{0}
\vspace{5mm}
\begin{center}
{\it Dedicated to Fa Yueh Wu on the occasion of his  80th birthday.}
\end{center}

\begin{abstract}

We define the notion of a spanning tree generating function (STGF) $\sum a_n z^n$, which gives the spanning tree constant when evaluated at $z=1,$ and gives the lattice Green function (LGF) when differentiated. By making use of known results for logarithmic Mahler measures of certain Laurent polynomials, and proving new results, we express the STGFs as hypergeometric functions for all regular two and three dimensional lattices (and one higher-dimensional lattice).  This gives closed form expressions for the spanning tree constants for all such lattices, which were previously largely unknown in all but one three-dimensional case. We show for all lattices that these can also be represented as Dirichlet $L$-series. Making the connection between spanning tree generating functions and lattice Green functions produces integral identities and hypergeometric connections, some of which appear to be new.
\end{abstract}

\section{Introduction}

There is a simple connection between spanning trees (ST) and lattice Green functions (LGFs). This connection has been used previously, as detailed below, but has not previously been systematically exploited.
It may be used to derive the ST growth constant from the LGF, and as a byproduct, we derive certain integral identities involving complete elliptic integrals of the first kind,
and connections between different hypergeometric functions, some of which appear to be new.

In the next section we define spanning trees and spanning tree generating functions (STGFs) and briefly review their properties, particularly those relevant to this study.
In the following section we define and provide relevant reviews of properties of lattice Green functions. In the fourth section we define logarithmic Mahler measures, and give some results needed in subsequent derivations. In section 5 we exploit the connection between LGFs and STGFs. In particular, the integrals defining the STGFs are readily seen to be expressible in terms of logarithmic Mahler measures. From the existing literature on Mahler measures \cite{R09, R12}, plus significant extensions given here for the simple-cubic case, all of the STGFs for the standard two- and three-dimensional lattices are shown to be expressible in terms of hypergeometric functions\footnote{After the completion of this work we became aware of the recent paper of G S Joyce \cite{J12}, who studied the problem on the face-centred cubic lattice and obtained results in agreement with ours, as well as several additional results on related problems}. We also express these constants in terms of Dirichlet $L$-series. In Section 6 we use these results to produce integral and hypergeometric identities. Section 7 briefly outlines the connection between spanning trees, dimer coverings and the Ising model.

\section{Spanning trees on a lattice}
A {\it spanning tree} on a graph $G$  is a loop-free connected graph connecting all sites of  $G.$ The number of spanning trees on a graph $G$ we denote $T_G.$ For a class of graphs called {\it recursive}\footnote{A class of graphs is recursive if it can be constructed by sequential addition of a given subgraph. Thus regular lattices are recursive.}, which includes regular lattices, Shrock and Wu \cite{SW00} have proved that the number of spanning trees grows exponentially with the number of sites of the lattice. That is to say, for a regular lattice $\mathcal {L}$ of $N$ sites, the number of spanning trees on $\mathcal {L},$ denoted $T_{\mathcal {L}}(N)$ grows like $e^{\lambda N}$ for $N$ large. The limit $\lim_{N \to \infty} \frac{1}{N} \log T_{\mathcal {L}}(N) = \lambda_{\mathcal {L}}$ exists, is greater than zero, and depends on the lattice $\mathcal {L}.$ We call this limit $ \lambda_{\mathcal {L}}$ the {\it  spanning tree constant.}  Lyons \cite{L05} refers to it as the {\em tree entropy.}

The exponential growth with the number of sites is not the case for all graphs. For example, as pointed out in \cite{SW00}, for the linear chain of $N$ sites there is but one spanning tree, whereas for the complete graph of $N$ sites the number of spanning trees grows faster than exponentially.

There are several ways of calculating the number of spanning trees for a given graph. The standard graph-theoretical method is by construction of the Laplacian matrix, see, for example \cite{B93}.
Spanning trees can also be related to a special value of the Tutte polynomial.

As pointed out by Wu \cite{W77}, if the underlying graph is periodic, that is to say a lattice, then following Fortuin and Kasteleyn \cite{FK} $T_{\mathcal {L}}(N)$ can be expressed as the partition function of a lattice model. This representation then allows one to calculate $T_{\mathcal {L}}(N)$ as the partition function of an ice-type model on a related lattice \cite{BKW76}. In this correspondence, the partition function is evaluated as a Pfaffian, and the spanning tree constant is given as a $d$-dimensional integral.
This integral may be written as $$\lambda_{\mathcal L} = \log{q} + \frac{1}{(2\pi)^d}\int_{-\pi}^{\pi} dk_1 \cdots  \int_{-\pi}^{\pi} dk_d \log(1 - \Lambda({\mathcal L})),$$ where $q$ is now the {\it co-ordination number} of the lattice\footnote{Not to be confused with the number of states of the Potts model} and $\Lambda({\mathcal L})$ is the {\it structure function} of the lattice. The co-ordination number is just the number of nearest neighbours per lattice site, while the structure function is the Fourier transform of the discrete step probability function. For example, for the $d$-dimensional hypercubic lattice ${\mathcal L}_d$, one has $q=2d$ and $$\Lambda({\mathcal L}_d) = \frac{1}{d}(\cos k_1 + \cos k_2 + \cdots + \cos k_d).$$
The above integral expression for the ST constant has subsequently been independently derived by graph theorists, first by Burton and Pemantle \cite{BP93}, and in a slightly simpler form by Lyons \cite{L05}.

There is a considerable body of literature on the problem of evaluating $\lambda_{\mathcal L}$ for a variety of lattices $\mathcal L.$ In two dimensions this can usually be done exactly, but not generally  in higher dimension (the hyper-body-centred cubic lattice is an exception to this remark).

In two dimensions the results for the square, triangular, honeycomb were first given by Wu in \cite{W77} and by Shrock and Wu \cite{SW00} for the kagome lattice. They are:
\begin{eqnarray}
\lambda_{sq}&=& \frac{1}{4\pi^2}\int_{-\pi}^{\pi} \int_{-\pi}^{\pi} dk_1 dk_2 \log[4-2(\cos{k_1} + \cos{k_2})]\\ \nonumber
 &=& \frac{4}{\pi}(1 - \frac{1}{3^2} + \frac{1}{5^2} - \frac{1}{7^2} + \ldots) = \frac{4G}{\pi} = 1.166243616\ldots \\ \nonumber
\lambda_{tri}&=& \frac{1}{4\pi^2}\int_{-\pi}^{\pi} \int_{-\pi}^{\pi} dk_1 dk_2 \log[6-2(\cos{k_1} + \cos{k_2}+\cos(k_1+k_2))]\\ \nonumber
 &=& \frac{3\sqrt{3}}{\pi}(1 - \frac{1}{5^2} + \frac{1}{7^2} - \frac{1}{11^2} + \frac{1}{13^2} + \ldots)  = 1.61532973\ldots \\ \nonumber
 \lambda_{honey}&=&\lambda_{tri}/2 \\ \nonumber
 \lambda_{kagome}&=&(\lambda_{tri}+ \log{6})/3. \\ \nonumber
 \end{eqnarray}

We remark that these results can also be expressed in terms of primitive Dirichlet $L$-series \cite{We}. In particular,
\begin{eqnarray}
\lambda_{sq}&=&  \frac{4}{\pi}L_{-4}(2)  \\ \nonumber
\lambda_{tri}&=& \frac{15\sqrt{3}}{4\pi}L_{-3}(2)  \\ \nonumber
 \lambda_{honey}&=&\frac{15\sqrt{3}}{8\pi}L_{-3}(2)\\ \nonumber
 \lambda_{kagome}&=&\left (\frac{5\sqrt{3}}{4\pi}L_{-3}(2)+ \frac{1}{3}\log{6}\right ). \\ \nonumber
 \end{eqnarray}

 In the above equation, $G$ is Catalan's constant; the result for the honeycomb lattice is a consequence of a theorem \cite{SW00} linking $\lambda_{\mathcal L}$ on a given lattice ${\mathcal L}$ to $\lambda_{\mathcal L*}$ on the dual lattice $ {\mathcal L*}.$ A variety of other two-dimensional lattices have also been considered in \cite{SW00} (as well as some lattices of higher dimension), while in \cite{CW06} it is pointed out that there is more than one way to define a lattice, which gives rise to a range of different integral expressions for the spanning tree constant. In this way Chang and Wang found a number of integral identities, by virtue of different choices of integrand, which all evaluate to the spanning tree constant for the lattice in question. In \cite{C08} some non-regular two-dimensional lattices are considered, and a systematic treatment of such latices is given, including the evaluation -- in most cases numerical rather than exact -- of the spanning tree constant for a wide variety of such lattices.

In what follows, it will turn out to be useful to generalise the integral and form what we call a {\it spanning tree generating function} (STGF). We define it as
\begin{equation}
\label{stgf}
T_{\mathcal L}(z) \equiv \log{q} + \frac{1}{(2\pi)^d}\int_{-\pi}^{\pi} dk_1 \cdots  \int_{-\pi}^{\pi} dk_d \log[1/z - \Lambda({\vec k})].
\end{equation}
Clearly, the spanning tree constant $\lambda_{\mathcal L}=T_{\mathcal L}(1).$
It immediately follows that
\begin{equation}
\label{stdif}
-z\frac{dT_{\mathcal L}(z) }{dz} =  \frac{1}{(2\pi)^d}\int_{-\pi}^{\pi} dk_1 \cdots  \int_{-\pi}^{\pi} dk_d \frac{1}{1 - z\Lambda({\vec k})}=P_{\mathcal L}(0,z).
\end{equation}

The integral in eqn(\ref{stdif}) will be recognised as the lattice Green function of lattice ${\mathcal L}.$

For some lattices, such as the honeycomb and diamond lattices, the LGF is given by
\begin{equation}
\label{lgfh}
 P_{\mathcal L}(0,z)
=\frac{1}{(2\pi)^d}\int_{-\pi}^{\pi} dk_1 \cdots  \int_{-\pi}^{\pi} dk_d \frac{1}{1 - z^2\Lambda({\vec k})}.
\end{equation}
In that case the STGF is
\begin{equation}
\label{stgfh}
T_{\mathcal L}(z) = \log{q} + \frac{1}{2}\frac{1}{(2\pi)^d}\int_{-\pi}^{\pi} dk_1 \cdots  \int_{-\pi}^{\pi} dk_d \log[1/z^2 - \Lambda({\vec k})],
\end{equation}
and eqn (\ref{stdif}) still holds.

This generalisation is not new. It appears to have first been used by Rosengren \cite{R87}, who obtained an expression for the spanning tree constant on the simple cubic lattice by making use of known recurrences for the LGF of the simple cubic lattice, and integrating these. More recently, Joyce, Delves and Zucker \cite{JDZ98}, inspired by an integral occurring in work of Baxter and Bazhanov (unpublished), which in structure is of the form of eqn(\ref{stgf}), instead investigated the derivative function, which gives rise to an integral of LGF type, as in eqn(\ref{stdif}). In \cite{J01} Joyce studied integrals equivalent to the STGFs, with a view to obtaining accurate estimates of the spanning tree constants. He evaluted the STGF for the body-centred cubic lattice in terms of hypergeometric functions. He also evaluated the spanning tree constants of the three standard three-dimensional lattices to extraordinary accuracy, nearly 200 decimal digits. Further, in 2005, Glasser and Lamb \cite{GL05} considered the lattice spanning tree entropy for various two-dimensional lattices, and used it to derive some anisotropic triangular lattice Green functions by differentiation.

However here we exploit this connection in a systematic manner. The existence of known exact results for some lattice Green functions can now be integrated and provide new, simpler, representations for the spanning tree generating function, as well as new integral identities.
More precisely, we have
\begin{equation}\label{tree}
T_{\mathcal L}(z) = \log{q} - \int  \frac{P(0,z)}{z} dz.
\end{equation}

Note first that we have restored the constant of integration, $\log{q},$ lost in the differentiation. Secondly, when the indefinite integral can be evaluated, and the STGF found, the spanning tree constant can be found by evaluating the STGF at $z = 1.$  In most cases, particularly for lattices of dimension greater than 2, the integral cannot be performed. In that case we can still get an expression for the spanning tree constant. First, note that, by definition, the LGF can be written $P(0,z) = 1 + {\rm O}(z).$ The integrand of (\ref{tree}) therefore has a simple pole at the origin, which contributes a term $\log{z}$ to the STGF. This of course vanishes when evaluating the STGF at $z=1,$ while all higher order terms vanish at $z=0.$ So if we artificially remove this pole at the origin of the integrand and integrate $\frac{P(0,z)-1}{z} $
between 0 and 1, we will obtain the spanning tree constant. That is to say,

\begin{equation}\label{treeconst}
\lambda_{\mathcal L} = \log{q} - \int_0^1  \frac{P(0,z) - 1}{z} dz.
\end{equation}
This result has been previously obtained by Lyons \cite{L05}.

In the next section we review some results for lattice Green functions which will be needed in our subsequent development of spanning tree generating functions.

\section{Lattice Green functions}

 For a translationally invariant walk on a $d$-dimensional periodic Bravais lattice, a natural question to ask is the probability that a walker starting at the origin will be at position ${\vec l}$ after $n$ steps.
 The probability generating function is known as the Lattice Green Function (LGF).
 It is
 \begin{equation}
\label{lgf}
P({\vec l};z)=\frac{1}{(2\pi)^d}\int_{-\pi}^{\pi} \cdots  \int_{-\pi}^{\pi}
 \frac{\exp(-i{\vec l}.{\vec k})d^d{\vec k}}{1-z\Lambda({\vec k})} .
 \end{equation}
 So that  $[z^n]P({\vec l};z)$ is the probability that a walker starting at the origin will be at ${\vec l}$ after $n$ steps.
 $\Lambda({\vec k})$ is the {\em structure function} of the lattice walk, as noted above.

The probability of return to the origin is $1 - 1/P({\vec 0};1).$
     $$P({\vec 0};1)=\frac{1}{(2\pi)^2}\int_{-\pi}^{\pi}  \int_{-\pi}^{\pi} \frac{dk_1\, dk_2}{1-\Lambda({\vec k})} $$
   Since $P({\vec 0};1)$ diverges for two-dimensional lattices, this leads to the well-known result that the probability of return to the origin in two dimensions is certain. In three dimensions these are the celebrated Watson integrals. The history of their development and evaluation has recently been authoritatively given by Zucker in \cite{Z11}.

    Of broader interest are LGFs defined by:
 \begin{equation}\label{lgf0}
P({\vec 0};z)=\frac{1}{(2\pi)^d}\int_{-\pi}^{\pi} \cdots  \int_{-\pi}^{\pi}
 \frac{d^d{\vec k}}{1-z\Lambda({\vec k})} .
 \end{equation}

For the regular two-dimensional lattices the structure functions are\footnote{Because the honeycomb lattice has two types of site, the expansion parameter $z$ in eqn (\ref{lgf0}) should be replaced by $z^2.$}:
\begin{align*}
 \Lambda({\vec k})_{honeycomb}=&\frac{1}{9}(1+4\cos^2{x}+4\cos{x}\cos{y}),\\
 \Lambda({\vec k})_{square}=&\frac{1}{2}(\cos{k_1}+\cos{k_2}),\\
 \Lambda({\vec k})_{triang}=&\frac{1}{3}(\cos{k_1}+\cos{k_2}+\cos{(k_1+k_2)}).
\end{align*}
   The corresponding LGFs are:
    \begin{equation}
    \label{hlgf}
     P({\vec 0};z)_{honey}=\frac{6\sqrt{3}}{\pi(3-z)\sqrt{(3-z)(1+z)}}{\bf K}(k)
    \end{equation}
    where $$k=\frac{4z^2}{(3-z)\sqrt{z(3-z)(1+z)}},$$ where ${\bf K}(z)$ is the complete elliptic integral of the first kind. For the square lattice, the result is remarkably simple,
     \begin{equation}
    \label{sqlgf}
    P({\vec 0};z)_{sq}=\frac{2}{\pi}{\bf K}(z),
    \end{equation}

while for the triangular lattice the LGF is:
 \begin{equation}
    \label{trlgf}
    P({\vec 0};z)_{tri}=\frac{6}{\pi z\sqrt{c}}{\bf K}(k')
    \end{equation}
where $c=(a+1)(b-1),$ and $$a=\frac{3}{z}+1-\sqrt{3+\frac{6}{z}}, \,\, {\rm and} \,\,  b=\frac{3}{z}+1+\sqrt{3+\frac{6}{z}} $$
 and $$k'=\sqrt{\frac{2(b-a)}{c}}.$$

 For the square lattice, we can also use the equivalent structure function
 $$\Lambda({\vec k})_{square}=\cos{k_1}\cos{k_2}. $$ as the square lattice can be considered as the two dimensional hyper-cubic lattice, which gives the first form of the structure function (above), or as the two-dimensional hyper-body-centred cubic lattice, giving the second form. While the integrands are clearly different, the integrals are equal, that is to say,
  \begin{equation}\label{lgf1}
P({\vec 0};z)_{square}=\frac{1}{(2\pi)^2}\int_{-\pi}^{\pi}   \int_{-\pi}^{\pi}
 \frac{dk_1 dk_2}{1-\frac{z}{2}(\cos{k_1}+\cos{k_2})} = \frac{1}{(2\pi)^2}\int_{-\pi}^{\pi}   \int_{-\pi}^{\pi}
 \frac{dk_1 dk_2}{1-z(\cos{k_1}\cos{k_2})}.
 \end{equation}
 Similarly, for the honeycomb lattice we can exploit the duality with the triangular lattice and write the structure function as $$\Lambda({\vec k})_{honey}=\frac{2}{3}(\frac{1}{2}+\Lambda({\vec k})_{tri}). $$
 It follows that
   \begin{eqnarray}\label{lgf2}
P({\vec 0};z)_{honey}&=&\frac{1}{(2\pi)^2}\int_{-\pi}^{\pi}   \int_{-\pi}^{\pi}
 \frac{dk_1 dk_2}{1-\frac{z^2}{3}(1+\frac{2}{3}[\cos{k_1}+\cos{k_2} + \cos(k_1+k_2)])} \\ \nonumber
 & = & \frac{1}{(2\pi)^2}\int_{-\pi}^{\pi}   \int_{-\pi}^{\pi}
 \frac{dk_1 dk_2}{1-\frac{z^2}{9}(1+4\cos^2{k_1}+4\cos{k_1}\cos{k_2})}. \nonumber
 \end{eqnarray}

 The LGFs for the three-dimensional lattices are also known. For the simple cubic lattice one has:
     $$P({\vec 0};z)=\frac{1}{(\pi)^3}\int_{0}^{\pi}  \int_{0}^{\pi} \int_{0}^{\pi}
 \frac{dk_1\, dk_2\, dk_3}{1-\frac{z}{3}(\cos{k_1}+\cos{k_2} + \cos{k_3})} $$
 Joyce \cite{J98} showed that this could be expressed as
 $$P({\vec 0};z)=\frac{1-9\xi^4}{(1-\xi)^3(1+3\xi)}\left [ \frac{2}{\pi}{\bf K}(k_1)\right ]^2;$$
  where $$k_1^2=\frac{16\xi^3}{(1-\xi)^3(1+3\xi)}; $$  with $$ \xi=(1+\sqrt{1-z^2})^{-1/2}(1-\sqrt{1-z^2/9})^{1/2}.$$

     For the body-centred cubic lattice one has:
     $$P({\vec 0};z)=\frac{1}{(\pi)^3}\int_{0}^{\pi}  \int_{0}^{\pi} \int_{0}^{\pi}
 \frac{dk_1\, dk_2\, dk_3}{1-z(\cos{k_1}\cos{k_2}\cos{k_3})}. $$
 Maradudin {\em et al.} \cite{MMW60} showed that this could be expressed as
 $$P({\vec 0};z)=\left [ \frac{2}{\pi}{\bf K}(k_2)\right ]^2$$
 where $$k_2^2=\frac{1}{2} - \frac{1}{2}\sqrt{1-z^2}.$$

     For the face-centred cubic lattice one has:
     $$P({\vec 0};z)=\frac{1}{(\pi)^3}\int_{0}^{\pi}  \int_{0}^{\pi} \int_{0}^{\pi}
  \frac{dk_1\, dk_2\, dk_3}{1-\frac{z}{3}(c_1c_2+c_1c_3+c_2c_3)} $$
  where $c_i=\cos{k_i}.$
   Joyce \cite{J98}) showed that this could be expressed as
 $$P({\vec 0};z)=\frac{(1+3\xi^2)^2}{(1-\xi)^3(1+3\xi)}\left [ \frac{2}{\pi}{\bf K}(k_3)\right ]^2;$$
 where $$k_3^2=\frac{16\xi^3}{(1-\xi)^3(1+3\xi)};  $$ and $$ \xi=(1+\sqrt{1-z})^{-1}(-1+\sqrt{1+3z}).$$

 Finally, for the diamond lattice one has:
  $$P({\vec 0};z)=\frac{1}{(\pi)^3}\int_{0}^{\pi}  \int_{0}^{\pi} \int_{0}^{\pi}
  \frac{dk_1\, dk_2\, dk_3}{1-\frac{z^2}{4}(1+c_1c_2+c_1c_3+c_2c_3)} ,$$
  which Joyce \cite{J73} pointed out gives
  $$P({\vec 0};z)=\frac{4}{\pi^2}{\bf K}(k_+){\bf K}(k_-);$$
 where $$k_{\pm}^2=\frac{1}{2} \pm \frac{1}{4}z^2(4-z^2)^{(1/2)} - \frac{1}{4}(2-z^2)(1-z^2)^{(1/2)}.$$

 The derivation of these results and some history and erudite discussion can be conveniently found in the book by Hughes \cite{H95}.


\section{Mahler measures}
The (logarithmic) Mahler measure of an $n$-variable Laurent polynomial is usually defined by
$$\m(P(z_1, \ldots , z_n)) := \int_0^1 \cdots \int_0^1 \log |P(\e^{2\pi \i \theta_1}, \ldots, \e^{2\pi \i \theta_n})| d\theta_1 \ldots d\theta_n. $$

In the two-variable case, we will be interested in the following polynomials, which we denote as shown:
\begin{eqnarray}\label{twom}
m(k)&:=&\m\left(k+x+\frac{1}{x} + y + \frac{1}{y}\right), \\ \nonumber
n(k)&:=&\m(x^3+y^3+1-kxy), \\ \nonumber
g(k)&:=&\m((1+x)(1+y)(x+y)-kxy). \\ \nonumber
\end{eqnarray}

The Mahler measures above can be represented in terms of hypergeometric functions.  The generalized hypergeometric function
is defined by
\begin{equation*}
{_pF_q}\left(\substack{a_1,\dots,a_p\\b_1,\dots,b_q};z\right)=\sum_{n=0}^{\infty}\frac{(a_1)_n\dots(a_p)_n}{(b_1)_n\dots(b_q)_n}\frac{z^n}{n!},
\end{equation*}
where $(a)_n=\Gamma(a+n)/\Gamma(a)$.  It was proved in \cite{RV} and \cite{LR} that
\begin{eqnarray}\label{twohyper}
m(k)&:=& \Re \left ( \log(k) - \frac{2}{k^2} {_4F_3}\left (\substack {\frac{3}{2},\frac{3}{2},1,1;\\2,2,2}\frac{16}{k^2}\right ) \right ), \\ \nonumber
n(k)&:=&\Re \left ( \log(k) - \frac{2}{k^3} {_4F_3}\left (\substack {\frac{5}{3},\frac{4}{3},1,1;\\2,2,2}\frac{27}{k^3}\right ) \right ), \\ \nonumber
g(k)&:=&\frac{1}{3} \Re \left ( \log\left ( \frac{(4+k)(k-2)^4}{k^2} \right )  -    \frac{2k^2}{(4+k)^3} {_4F_3}\left ( \substack{ \frac{5}{3},\frac{4}{3},1,1;\\2,2,2}\frac{27k^2}{(4+k)^3}\right )\right.\\ \nonumber
 &&\qquad\qquad\left.-  \frac{8k}{(k-2)^3} {_4F_3}\left (\substack {\frac{5}{3},\frac{4}{3},1,1;\\2,2,2}\frac{27k}{(k-2)^3}\right )\right ). \\ \nonumber
\end{eqnarray}
The derivation of the formula for $g(k)$ requires a modular expansion due to Stienstra \cite{S05}, as mentioned in \cite{LR}.

In the three-variable case we will be interested in the following polynomials:
\begin{eqnarray}\label{threem}
p(k)&:=&\m\left( \sqrt{k}+\left (x+\frac{1}{x}\right )\left (y + \frac{1}{y}\right )\left (z + \frac{1}{z}\right )\right ), \\ \nonumber
s(k)&:=&\m\left (k+x+\frac{1}{x} + y + \frac{1}{y}+ z + \frac{1}{z}\right ), \\ \nonumber
f(k)&:=&\m \left (4-k+\left (x+\frac{1}{x}\right )\left (y + \frac{1}{y}\right ) + \left (x+\frac{1}{x}\right )\left (z + \frac{1}{z}\right )+\left (y + \frac{1}{y}\right )\left (z + \frac{1}{z}\right )\right ). \\ \nonumber
\end{eqnarray}
In \cite{R09} one of us proved that for  $u$ sufficiently large
\begin{eqnarray}\label{threehyper}
p(u)&:=&\frac{1}{2}f_2(u), \\ \nonumber
s(3(u+u^{-1}))&:=&\frac{1}{20}f_4\left ( \frac{9(3+u^2)^4}{u^6} \right ) + \frac{3}{20}f_4\left ( \frac{9(3+u^{-2})^4}{u^{-6}} \right ), \\ \nonumber
f(u)&:=&-\frac{1}{15}f_3\left ( \frac{(16-u)^3}{u^2} \right ) + \frac{8}{15}f_3\left (- \frac{(4-u)^3}{u} \right ). \\ \nonumber
\end{eqnarray}
where, for $|u|$ sufficiently large,
\begin{eqnarray}\label{fs}
f_2(u)&=&\Re \left ( \log(u) - \frac{8}{u} {_5F_4}\left (\substack {\frac{3}{2},\frac{3}{2},\frac{3}{2},1,1;\\2,2,2,2} \frac{64}{u}\right ) \right ),
\\  \nonumber
f_3(u) &=&\Re \left ( \log(u) - \frac{12}{u} {_5F_4}\left (\substack {\frac{5}{3},\frac{3}{2},\frac{4}{3},1,1;\\2,2,2,2}\frac{108}{u}\right ) \right ),
 \\ \nonumber
f_4(u) &=&\Re \left ( \log(u) - \frac{24}{u} {_5F_4}\left (\substack {\frac{5}{4},\frac{3}{2},\frac{7}{4},1,1;\\2,2,2,2} \frac{256}{u}\right ) \right ).
 \\ \nonumber
\end{eqnarray}

For our purposes here, we point out that

\begin{eqnarray}\label{threes}
s(1/u) &=& \Re \left [ -\log(u) -\frac{1}{2} \sum_{n=1}^\infty \frac{u^{2n}}{n} {\binom{2n}{n}} \sum_{k=0}^n {\binom{2k}{k}} {\binom{n}{k}}^2 \right ] \,\,\, {\rm for} \,\, |u| <\frac{1}{6}, \\ \nonumber
f(1/u) &=& \Re \left [ -\log(u) - \sum_{n=1}^\infty \frac{u^{n}}{n}  \sum_{k=0}^n {\binom{2n-2k}{n-k}}{\binom{2k}{k}} {\binom{n}{k}}^2 \right ]  \,\,\, {\rm for} \,\, |u| < \frac{1}{16}. \\ \nonumber
\end{eqnarray}

\section{Connections between spanning tree generating functions, Mahler measures and lattice Green functions.}
From equation (\ref{stdif}), one can clearly calculate the spanning tree generating function from the lattice Green function (\ref{lgf}), up to a known additive constant, notably $\log q,$ where $q$ is the co-ordination number of the lattice. Furthermore, from the results given above for Mahler measures, we can in principle express the STGFs for all lattices in terms of Mahler measures, and hence in terms hypergeometric functions.

\subsection{Two-dimensional lattices}

For example, for the square lattice, from eqns (\ref{twom}) and (\ref{twohyper}) one  has
\begin{eqnarray}\label{treesq}
T_{sq}(z) &=& \log{4} + \frac{1}{4\pi^2} \int_{-\pi}^{\pi} \int_{-\pi}^{\pi} \log\left [\frac{1}{z}-   \frac{1}{2} (\cos{k_1} + \cos {k_2}) \right ] dk_1\cdot dk_2 \\ \nonumber
&=& \log{4} + \frac{1}{4\pi^2} \int_{-\pi}^{\pi} \int_{-\pi}^{\pi} \log\left |\frac{1}{z}-   \frac{1}{2} (\cos{k_1} + \cos {k_2}) \right | dk_1\cdot dk_2 \\ \nonumber
& = & m(4/z) \\ \nonumber
& =& \log{4} - \log{z} - \frac{z^2}{8} {_4F_3}\left (\substack{1,1,\frac{3}{2},\frac{3}{2}\\2,2,2};z^2\right ). \\ \nonumber
\end{eqnarray}
The second equality follows since the function inside the logarithm is greater than or equal to zero for $0 \le z \le 1,$ allowing us to introduce the absolute value sign inside the logarithm, thus turning the expression into a logarithmic Mahler measure. If we substitute $x=e^{2\pi i k_1}$ and $y=e^{2\pi i k_2}$, then the third equality immediately follows.

In this case, we can also obtain the result by direct integration of the LGF, since for this lattice $P({\vec 0};z)=\frac{2}{\pi}{\bf K}(z),$ so from eqn (\ref{tree})

\begin{equation}\label{treesq}
T_{sq}(z) = \log{4} -  \frac{2}{\pi} \int  \frac{{\bf K}(z)}{z} dz = \log{4} - \log{z} - \frac{z^2}{8} {_4F_3}\left (\substack{1,1,\frac{3}{2},\frac{3}{2}\\2,2,2};z^2\right ).
\end{equation}
From \cite{R12} the r.h.s can be further simplified to $$\Re \left (z \cdot {_3F_2}\left (\substack{ \frac{1}{2},\frac{1}{2},\frac{1}{2}\\1,\frac{3}{2}};\frac{1}{z^2}\right ) \right ).$$
It immediately follows that,
\begin{eqnarray}\label{lamsq}
\lambda_{sq} & = & \log{4} - \int_0^1  \frac{\frac{2}{\pi}{\bf K}(z) - 1}{z} dz =  \log{4}  - \frac{1}{8} {_4F_3}\left (\substack{1,1,\frac{3}{2},\frac{3}{2}\\2,2,2};1\right ) \\ \nonumber
& =& {_3F_2}\left (\substack{\frac{1}{2},\frac{1}{2},\frac{1}{2}\\1,\frac{3}{2}};1\right ) =  \frac{4G}{\pi}. \nonumber
\end{eqnarray}

\noindent
For the hexagonal lattice,


\begin{align}\label{treehex}
T_{hex}(z)=&\log3+\frac{1}{8\pi^2}\int_{-\pi}^{\pi}\int_{-\pi}^{\pi}\log\left[\frac{1}{z^2}-\frac{1}{9}\left(1+4\cos^2k_1+4\cos k_1 \cos k_2\right)\right]dk_1 dk_2\\
=&\frac{1}{8\pi^2}\int_{-\pi}^{\pi}\int_{-\pi}^{\pi}\log\left[\left(\frac{9}{z^2}-1\right)-\left(4\cos^2k_1+4\cos k_1 \cos k_2\right)\right]dk_1 dk_2
\end{align}
Let us set
\begin{equation*}
k=\frac{9}{z^2}-1.
\end{equation*}
The function inside the logarithm is greater than or equal to zero if $k\ge8$, or equivalently if $-1\le z\le1$.  This allows us to introduce an absolute value sign inside the logarithm.
We obtain
\begin{align*}
T_{hex}(z)=&\frac{1}{8\pi^2}\int_{-\pi}^{\pi}\int_{-\pi}^{\pi}\log\left|k-\left(4\cos^2k_1+4\cos k_1 \cos k_2\right)\right|dk_1 dk_2\\
=&\frac{1}{2}\int_{0}^{1}\int_{0}^{1}\log\left|k-4\cos\left(2\pi k_1\right)\left(\cos(2\pi k_1)+\cos \left(2\pi k_2\right)\right)\right|dk_1 dk_2\\
=&\frac{1}{2}\int_{0}^{1}\int_{0}^{1}\log\left|k-8\cos\left(4\pi k_1\right)\cos(2\pi (k_1-k_2))\cos \left(2\pi(k_1+ k_2)\right)\right|dk_1 dk_2.
\end{align*}
If $x=e^{2\pi i k_1}$ and $y=e^{2\pi i k_2}$, then
\begin{align*}
T_{hex}(z)=&\frac{1}{2}\int_{0}^{1}\int_{0}^{1}\log\left|k-\left(x^2+x^{-2}\right)\left(\frac{x}{y}+\frac{y}{x}\right)\left(x~y+\frac{1}{x~y}\right)\right|dk_1 dk_2\\
=&\frac{1}{2}\m\left(k-\left(x^2+x^{-2}\right)\left(\frac{x}{y}+\frac{y}{x}\right)\left(x~y+\frac{1}{x~y}\right)\right).
\end{align*}
We are now dealing with a Mahler measure.  Let $y\rightarrow \frac{y}{x}$.  Then
\begin{align*}
T_{hex}(z)=&\frac{1}{2}\m\left(k-\left(x^2+\frac{1}{x^{2}}\right)\left(\frac{x^2}{y}+\frac{y}{x^2}\right)\left(y+\frac{1}{y}\right)\right),\\
=&\frac{1}{2}\m\left(\left(x^4+1\right)(y^2+1)\left(x^4+y^2\right)-k x^4 y^2\right).
\end{align*}
Finally let $(x^4,y^2)\rightarrow (x,y)$, and we arrive at
\begin{align*}
=&\frac{1}{2}\m\left(\left(1+x\right)\left(1+y\right)\left(x+y\right)-k x y\right)\\
=&\frac{1}{2}g(k).
\end{align*}
Therefore, for $-1\le z\le 1$, we conclude that
\begin{equation}
T_{hex}(z)=\frac{1}{2}g\left(\frac{9}{z^2}-1\right).
\end{equation}
When $z=1$, we obtain
\begin{equation*}
\lambda_{hex} =T_{hex}(1)=\frac{1}{2}g(8)=\frac{1}{2}\m\left((1+x)(1+y)(x+y)-8x y \right)=\frac{15\sqrt{3}}{8\pi}L_{-3}(2).
\end{equation*}
The evaluation of $g(8)$ is essentially due to Boyd, and a closely related calculation is described in \cite[p.~31]{RV}.  To recap Boyd's result, let us
use \eqref{twohyper} to obtain
\begin{align*}
g(8)=&\frac{5}{3}\log3-\frac{10}{81}{_4F_3}\left(\substack{\frac53,\frac43,1,1\\2,2,2};1\right)\\
=&\frac{5}{6}\m\left(x^3+y^3+1-3 xy\right).
\end{align*}
The Mahler measure $n(k):=\m\left(x^3+y^3+1-k x y\right)$ also reduces to ${_4F_3}$ functions \cite{LR}.
If $\omega=e^{2\pi i/3}$, then factoring the polynomial gives
\begin{align*}
=&\frac{5}{6}\m\left((x+y+1)(x+\omega^2 y+\omega)(x+\omega y+\omega^2)\right)\\
=&\frac{5}{2}\m\left(1+x+y\right),\\
=&\frac{15\sqrt{3}}{8\pi}L_{-3}(2).
\end{align*}
The second step follows from additivity of Mahler measures, plus elementary changes of variable.  The final step requires an evaluation due to Smyth \cite{B81}.

\noindent
Similarly, for the triangular lattice,
\begin{equation}\label{treetr}
T_{tri}(z) = \log{6} -  \frac{6}{\pi} \int \frac{\sqrt{3}{\bf K}(k)}{ z(3-z)\sqrt{(3-z)(1+z)}} dz
    \end{equation}
    where $k=\frac{4z^2}{(3-z)\sqrt{z(3-z)(1+z)}},$
and from eqn (\ref{treeconst}),
\begin{eqnarray}\label{lamtrr}
\lambda_{tri} & = & \log{6} - \int_0^1 \left [\frac{6\sqrt{3}{\bf K}(k)}{\pi z^2 (3-z)\sqrt{(3-z)(1+z)}} - \frac{1}{z} \right ] dz \\ \nonumber
& = & \frac{3\sqrt{3}}{\pi}(1 - \frac{1}{5^2} + \frac{1}{7^2} - \frac{1}{11^2} + \frac{1}{13^2} + \ldots)= \frac{1}{\sqrt{3}}\left ( \pi - \frac{\Psi'(5/6)}{2\pi} \right )\\ \nonumber
&  =& 1.615329736\ldots \\ \nonumber
\end{eqnarray}

Now we would like to evaluate the function
\begin{equation*}
T_{tri}(z)=\log6+\frac{1}{4\pi^2}\int_{-\pi}^{\pi}\int_{-\pi}^{\pi}\log\left(\frac{1}{z}-\frac{1}{3}\left(\cos k_1+\cos k_2+\cos(k_1+k_2)\right)\right)d k_1 d k_2
\end{equation*}
Suppose that $0<z\le 1$.  Then we can insert an absolute value inside the logarithm, and we are again dealing with a Mahler measure.
\begin{align*}
T_{tri}(z)=&\log6+\frac{1}{4\pi^2}\int_{-\pi}^{\pi}\int_{-\pi}^{\pi}\log\left|\frac{1}{z}-\frac{1}{3}\left(\cos k_1+\cos k_2+\cos(k_1+k_2)\right)\right|d k_1 d k_2\\
=&\int_{0}^{1}\int_{0}^{1}\log\left|\frac{6}{z}+2\left(\cos(2\pi k_1)+\cos(2\pi k_2)-\cos(2\pi(k_1+k_2))\right)\right|d k_1 d k_2\\
=&\m\left(\frac{6}{z}+x+\frac{1}{x}+y+\frac{1}{y}-x y-\frac{1}{x y}\right).
\end{align*}
Now let $(x,y)\mapsto\left(x y,\frac{y}{x}\right)$, to obtain
\begin{align*}
T_{tri}(z)=&\m\left(\frac{6}{z}+x y+\frac{1}{x y}+\frac{y}{x}+\frac{x}{y}-y^2-\frac{1}{y^2}\right)\\
=&\m\left(\frac{6}{z}+2+\left(x+\frac{1}{x}\right)\left(y+\frac{1}{y}\right)-\left(y+\frac{1}{y}\right)^2\right)\\
=&\m\left(\frac{6}{z}+2+\left(y+\frac{1}{y}\right)\left(x+\frac{1}{x}-y-\frac{1}{y}\right)\right).
\end{align*}
Finally, we make the substitution $(x,y)\mapsto\left(x y,-\frac{y}{x}\right)$, and the formula becomes
\begin{align*}
T_{tri}(z)=&\m\left(\frac{6}{z}+2-\left(\frac{x}{y}+\frac{y}{x}\right)\left(x+\frac{1}{x}\right)\left(y+\frac{1}{y}\right)\right)\\
=&\m\left((x+1)(y+1)(x+y)-\left(\frac{6}{z}+2\right) x y\right)\\
=&g\left(2+\frac{6}{z}\right).
\end{align*}

In summary, we have the following result:
\begin{equation}
T_{tri}(z)=g\left(2+\frac{6}{z}\right),
\end{equation}
whenever $z\in(0,1]$.  When $z=1$ we obtain
\begin{equation*}
\lambda_{tri}=T_{tri}(1)=g(8)=\frac{15\sqrt{3}}{4\pi}L_{-3}(2),
\end{equation*}
after appealing to evaluation of $g(8)$ described above.
In \cite{GW05} Glasser and Wu give two other approaches to the evaluation of the spanning tree constant for the triangular lattice, including its expression in terms of the Clausen function, $$Cl_2(\theta) = \sum_{n=1}^\infty \frac{\sin{(n\theta)}}{n^2}.$$ It is $\lambda_{tri} =\frac{5}{\pi}Cl_2\left ( \frac{\pi}{3} \right ).$

From the connection between the LGFs, and the known expressions for their coefficients \cite{G10}, we have the following identites:
$$\sum_{m > 0} {2m \choose m}^2 \frac{1}{m} \left ( \frac{1}{4} \right )^{2m} = \log{16} - \frac{8G}{\pi}.$$
This is essentially due to Ramanujan. An equivalent version appears in \cite{Be89} on page 40. It can also be readily proved by expressing both sides in terms of hypergeometric functions.

An interesting equality between lattice sums arises from the honeycomb--triangular duality, $\lambda_{tri}/2=\lambda_{honey}$. It follows that
\begin{eqnarray*}
&&\log{\frac{2}{3}}+\sum_{n > 0} \frac{1}{n}\left (\frac{1}{9}\right )^n \sum_{j= 0}^n {2j \choose j}{n \choose j}^2 \\ \nonumber
&=&\sum_{n > 0}  \frac{1}{n} \left ( \frac{1}{6} \right )^{n} \sum_{j= 0}^n (-3)^{n-j}{n \choose j} \sum_{k= 0}^j {j \choose k}^2 {2k \choose k}.
\end{eqnarray*}

\subsection{Three-dimensional lattices}
For the $d$-dimensional hyper-body-centred cubic lattice, the spanning tree generating function is
\begin{eqnarray}
\label{hyperbcc}
T_d^{bcc}(z)&=&\log{2^d}+\frac{1}{\pi^d}\int_0^{\pi} dk_1 \cdots \int_0^{\pi} dk_d \log \left ( \frac{1}{z} - \cos(k_1)\cdot\cos(k_2)\cdot \cdots \cos(k_d)\right )\\ \nonumber
&=& \log{2^d}+\frac{1}{\pi^d}\int_0^{\pi} dk_1 \cdots \int_0^{\pi} dk_d \log \left (1 - z\cos(k_1) \cdot\cos(k_2)\cdot \cdots \cos(k_d)\right) - \log(z)\\ \nonumber
&=& d\log(2)-\log(z)-\frac{1}{2}\sum_{l=1}^\infty \frac{z^l}{l}\left ( \frac{(2l)!}{2^{2l}(l!)^2} \right )^d \\ \nonumber
&=&d\log(2) - \log(z)- \frac{z^2}{2^{d+1}}  {_{d+2}F_{d+1}} \left (\substack {1,1\frac{3}{2},\ldots,\frac{3}{2}\\2,2,\ldots,2};z^2\right )
\end{eqnarray}
So the spanning tree constant $\lambda_c$ is
$$\lambda_c^{bcc}(d) = d\log(2)-\frac{1}{2}\sum_{l=1}^\infty \frac{1}{l}{2l \choose l}^d\left (\frac{1}{4^d}\right )^l=d\log(2) - \frac{1}{2^{d+1}}  {_{d+2}F_{d+1}} \left (\substack {1,1\frac{3}{2},\ldots,\frac{3}{2}\\2,2,\ldots,2};1\right ),$$ a result first given by Chang and Shrock \cite{CS06}.
We give the results for $d = 2, \, 3, \,4$ below, calculated to 50 significant digits almost instantaneously by Maple.
\begin{eqnarray}
\lambda_c^{bcc}(2) &=& 1.1662436161232751205535378258735796754562646159433 = 4G/\pi \\ \nonumber
\lambda_c^{bcc}(3) &=& 1.9901914182719407717105190854333649929453600034709 \\ \nonumber
\lambda_c^{bcc}(4) &=& 2.7329575354773621769814874419935610996620191115587. \nonumber
\end{eqnarray}
The result for the three-dimensional case also follows immediately from the Mahler measure identification given by eqn. (\ref{threehyper}), so that $$\lambda_c^{bcc}(3) = p(64) = \frac{1}{2}f_2(64).$$

Here $G = 0.915965594177219\ldots$ is Catalan's constant, and the result for $d=2$ is well known as the square lattice spanning tree  constant. For $d=3$ and $d=4$ the results lie outside the upper and lower bounds given by Felker and Lyons \cite{FL03}. In both cases their upper bounds lie below the correct result.

Because of the factorisation property peculiar to this lattice, the result is so simple that it has been derived without reference to Mahler measures. However it could have been derived that way, as   the relevant Mahler measure is given by equation (\ref{threem}) above. The value of the spanning tree constant is obtained by evaluating the STGF at $z=1,$ and this evaluation is given by Samart in terms of $L$ functions \cite{S12} as $$\lambda_c^{bcc}(3) = \frac{64}{\pi^3}L(\eta(4\tau)^6,3).$$  Here $\eta(\tau)$ is the Dedekind eta function.

\subsection{Hyper simple cubic lattice}
The bcc lattice STGF above was the simplest to obtain. The corresponding s.c. result, as we show below, is the most difficult.
For the $d$-dimensional simple cubic lattice, the spanning tree generating function is
\begin{eqnarray}
\label{hyperbcc} \label{hypersc}
T_d^{sc}(z)&=&\log(2d)+\frac{1}{\pi^d}\int_0^{\pi} d\theta_1 \cdots \int_0^{\pi} d\theta_d \log \left ( \frac{1}{z} - \frac{1}{d}[\cos(\theta_1) + \cdots + \cos(\theta_d)]\right )\\ \nonumber
&=&\log(2d)+\frac{1}{\pi^d}\int_0^{\pi} d\theta_1 \cdots \int_0^{\pi} d\theta_d \log \left (1 - \frac{z}{d}[\cos(\theta_1) + \cdots  + \cos(\theta_d)]\right) - \log(z)\\ \nonumber
&=& \log(2d)-\log(z)-\frac{1}{2}\sum_{l=1}^\infty \frac{1}{l} \left( \frac{z}{4d^2}\right )^l a_l^{(d)}
\end{eqnarray}
where the corresponding lattice Green function is
$$P_d({\vec 0};z)=\sum_{l=0}^\infty a_l^{(d)} z^l.$$

In particular, one has (see \cite{G09} for the 4-dimensional result):
$$ a_l^{(2)}= {2l \choose l} \sum_{j=0}^l {l \choose j}^2 = {2l \choose l}^2$$
$$ a_l^{(3)}= {2l \choose l} \sum_{j=0}^l {l \choose j}^2 {2j \choose j} = {2l \choose l} {_3F_2}\left (\substack {\frac{1}{2},-l,-l\\1,1};4\right )$$
$$ a_l^{(4)}= {2l \choose l} \sum_{j=0}^l {l \choose j}^2 {2j \choose j}{2l - 2j \choose l -j} = {2l \choose l}^2 {_4F_3}\left (\substack {\frac{1}{2},-l,-l,-l\\1,1, \frac{1}{2}-l};1\right )$$


It is a straightforward matter to sum the series (\ref{hypersc}), with $z$ set to $1,$ to any reasonably desired accuracy with one's favourite computer algebra package, and observing that the error made in stopping the summation after $n$ terms decreases exponentially with $n,$ we readily found
$$\lambda_c^{sc}(2) =T_2(1) = 1.166243616123\ldots,$$
$$\lambda_c^{sc}(3) =T_3(1) = 1.673389302970\ldots $$
$$\lambda_c^{sc}(4) =T_4(1) = 1.999707644517\ldots .$$

Joyce and Zucker \cite{JZ01} have made a thorough study of this problem for the case $z=1,$ and have reported the value of related integrals for the case $z=1$ to some 55 significant digits. The result in the two-dimensional case is exactly known, as given above, and we show below that we can evaluate the spanning tree generating function, and constant, exactly in the three-dimensional case also. In \cite{TW00} Tzeng and Wu have also studied the problem of spanning trees on hybercubic lattices, as well as on non-orientable surfaces.

From eqn. (\ref{threem}) we see that the simple cubic lattice STGF can be readily expressed as the logarithmic Mahler measure $$s \left ( - \frac{6}{z} \right ). $$
This is, in terms of a series expansion,
\begin{equation}\label{scstgf}
T_{sc}(z) = \log(6)-\frac{1}{2}\sum_{l=1}^\infty \frac{1}{l} \left( \frac{z}{36}\right )^l {2l \choose l} \sum_{j=0}^l {l \choose j}^2 {2j \choose j}.
\end{equation}
Now we have from \cite{R09}
\begin{eqnarray}\label{scmm}
& &\sum_{l=1}^\infty \frac{1}{l} \left( \frac{u}{9(1+u)^2}\right )^l {2l \choose l} \sum_{j=0}^l {l \choose j}^2 {2j \choose j} = \frac{2}{5}\log \left ( \frac{27(1+u)^5}{(3+u)^3(1+3u)} \right ) +\\ \nonumber
& & \frac{4u^3}{5(3+u)^4} {_5F_4}\left (\substack {\frac{5}{4},\frac{3}{2},\frac{7}{4},1,1\\2,2,2,2};\frac{256u^3}{9(3+u)^4}\right ) + \frac{4u}{15(1+3u)^4} {_5F_4}\left (\substack { \frac{5}{4},\frac{3}{2},\frac{7}{4},1,1\\2,2,2,2};\frac{256u}{9(1+3u)^4}\right ).
\end{eqnarray}
Unfortunately the argument of the second hypergeometric function is 1 when $u=1/9,$ so this expansion is only valid for $u \in [0, 1/9],$ whereas we require $z \in [0,1]$ which corresponds to $u \in [0,1].$


That is to say, we would like to extend the formula (\ref{scmm})
to all values of $u\in[0,1]$. The identity is valid when $u$ lies in a neighborhood of zero, and it holds on the positive real axis if $u\in[0,\frac19]$.
 It fails when $u>\frac{1}{9}$, because the argument of the second hypergeometric function intersects a branch cut on the interval $[1,\infty)$.  It is easy to see that $\frac{256u}{9(1+3u)^4}=1$ when $u=\frac19$.  In general, we can analytically continue \eqref{scmm} along a ray from $u=0$, until we reach a value for which either $\frac{256u}{9(1+3u)^4}\in[1,\infty)$, or $\frac{256u^3}{9(3+u)^4}\in[1,\infty)$.  When we cross a branch cut, it is necessary to correct an identity with additional terms, which are usually related to Meijer $G$-functions.

For numerical purposes, let us note that the following formula
\begin{equation}\label{numerical identity}
\begin{split}
\sum_{n=1}^{\infty}\frac{1}{n}&\left(\frac{u}{9(1+u)^2}\right)^n{2n\choose n}\sum_{k=0}^{n}{2k\choose k}{n\choose k}^2\\
=&\frac{2}{5}\log\left(\frac{27(1+u)^5}{(3+u)^3(1+3u)}\right)+\frac{4u^3}{5(3+u)^4}
{_5F_4}\left(\substack{\frac{5}{4},\frac{3}{2},\frac{7}{4},1,1\\2,2,2,2};\frac{256u^3}{9(3+u)^4}\right)\\
&+\frac{1}{10}\int_{0}^{w}\frac{\left({_2F_1}\left(\substack{\frac14,\frac34\\1};t\right)\right)^2-1}{t(1-t)}(1-2t)dt,
\end{split}
\end{equation}
holds if
\begin{align*}
u=\frac{9r}{(1-r)(8+r)},&&w=\frac{r (8+r)^3}{\left(8+20 r-r^2\right)^2},
\end{align*}
for all $r\in[0,6\sqrt{2}-8]$.  This restriction on $r$ implies that $u\in[0,1]$ and $w\in[0,\frac{3 + 2 \sqrt{2}}{6}]\approx [0,0.97]$.  The integral on the right is analytic when $|w|<1$, so it is easy to see that formula \eqref{numerical identity} is valid on a somewhat larger domain than formula \eqref{scmm}.  Formula \eqref{numerical identity} is an intermediate step in the derivation of \eqref{scmm}, but is not actually stated in \cite{R09}.  The two identities coincide if $w\in[0,\frac12]$, because Clausen's identity allows us to perform the integration \cite{CTYZ}:
\begin{equation*}
\begin{split}
\int_{0}^{w}\frac{\left({_2F_1}\left(\substack{\frac14,\frac34\\1};t\right)\right)^2-1}{t(1-t)}(1-2t)dt=&\int_{0}^{w}\frac{{_3F_2}\left(\substack{\frac14,\frac12,\frac34\\1,1};4t(1-t)\right)-1}{t(1-t)}(1-2t)dt\\
=&\int_{0}^{4w(1-w)}\frac{{_3F_2}\left(\substack{\frac14,\frac12,\frac34\\1,1};t'\right)-1}{t'}d t'\\
=&\frac{3}{8}w(1-w){_5F_4}\left(\substack{\frac54,\frac32,\frac74\\1,1};4w(1-w)\right).
\end{split}
\end{equation*}
Since $4w(1-w)=\frac{256u}{9(1+3u)^4}$, we recover \eqref{scmm}.  Clausen's identity fails when $w>1/2$, so we need an additional method to simplify \eqref{numerical identity}.

We simplify \eqref{numerical identity} by using modular parameterizations for hypergeometric functions.  Consider the following $q$-series:
\begin{equation}
G(q):=\Re\left[-\log(q)+240\sum_{n=1}^{\infty}n^2\log(1-q^n)\right],
\end{equation}
and note that
\begin{equation}\label{G addition}
G(q)+G(-q)=9G(q^2)-4G(q^4).
\end{equation}
Formula \eqref{G addition} is proved in \cite[Thm.~2.3]{R09}.  Recall that $f_{2}(u)$ and $f_{4}(u)$ are defined in eqn (\ref{fs}), and define the sum
\begin{align}
g_{1}(u):=&\Re\left[\log(u)-\sum_{n=1}^{\infty}\frac{(1/u)^{2n}}{2n}{2n\choose n}\sum_{k=0}^{n}{2k\choose k}{n\choose k}^2\right].
\end{align}
Results of Bertin \cite{Bert} and Rogers \cite{R09}
show that if $|q|$ is sufficiently small, then
\begin{align}
g_1(t_1(q))=&-\frac{1}{60}G(q)+\frac{1}{30}G(q^2)-\frac{1}{20}G(q^3)+\frac{1}{10}G(q^6),\label{g1 in terms of G}\\
f_{2}(s_2(q))=&-\frac{1}{15}G(q)+\frac{4}{15}G(q^4),\label{f2 in terms of G}\\
f_{4}(s_4(q))=&-\frac{1}{3}G(q)+\frac{2}{3}G(q^2),\label{f4 in terms of G}
\end{align}
where
\begin{align*}
t_1(q)=&\left(\frac{\eta(q)\eta(q^6)}{\eta(q^2)\eta(q^3)}\right)^6+\left(\frac{\eta(q)\eta(q^6)}{\eta(q^2)\eta(q^3)}\right)^{-6}\\
s_2(q)=&\left(\frac{\eta^2(q^2)}{\eta(q)\eta(q^4)}\right)^{24},\\
s_4(q)=&\left(\frac{\eta(q^2)}{\eta(q)}\right)^{24}\left(16\frac{\eta^4(q)\eta^{8}(q^4)}{\eta^{12}(q^2)}+\frac{\eta^{12}(q^2)}{\eta^4(q)\eta^{8}(q^4)}\right)^4,
\end{align*}
and
\begin{equation*}
\eta(q)=q^{1/24}\prod_{n=1}^{\infty}(1-q^n)=\sum_{n=-\infty}^{\infty}(-1)^n q^{(6n+1)^2/24}.
\end{equation*}
We note that Rogers gave a slightly different version of \eqref{f2 in terms of G}, and that this form of the identity was first observed by Samart \cite{S12}.  Rodriguez-Villegas and Stienstra proved similar modular expansions for two-variable Mahler measures \cite{S05}, \cite{RV}.  Using formulas \eqref{G addition}, \eqref{g1 in terms of G}, \eqref{f2 in terms of G}, and \eqref{f4 in terms of G}, we deduce the following five term relation:
\begin{equation}\label{five term relation}
\begin{split}
g_1(t_1(q))=&-\frac{1}{4} f_2\left(s_2\left(-q\right)\right)  + \frac{9}{20} f_4\left(s_4\left(q^2\right)\right)+\frac{3}{20} f_4\left(s_4\left(q^3\right)\right)\\
& + \frac{1}{10}f_4\left(s_4\left(-q^2\right)\right) - \frac{1}{5} f_4\left(s_4\left(q^4\right)\right),
\end{split}
\end{equation}
which holds for $q$ in a neighborhood of zero.  There are many additional relations between the hypergeometric functions, but this is the simplest formula we were able to find which holds for $q\in\left[0,e^{-\pi\sqrt{2/3}}\right]$.  We require the identity to be valid on a region which includes $q=e^{-\pi\sqrt{2/3}}$, because the s.c. constant is given by $g_1(6)=g_1\left(t_1\left(e^{-\pi\sqrt{2/3}}\right)\right)$.  Formula \eqref{five term relation} holds in a neighborhood of $q=0$, and it can be extended to the desired region, because the arguments of the hypergeometric functions never intersect the line $[1,\infty)$.  For instance, we can use a computer to check that $\frac{64}{s2(-q)}\not\in[1,\infty)$ for all $q\in\left[0,e^{-\pi\sqrt{2/3}}\right]$.

Equation \eqref{five term relation} simplifies if we observe that the functions $\{t_1(q), s_2(-q), s_4(q^2),\dots\}$ are algebraically dependent.  To obtain explicit relations, we need modular equations from Ramanujan's notebooks \cite{Be89}.  Suppose that $\alpha$ and $\beta$ are given by
\begin{align*}
\alpha:=16\frac{\eta^8(q)\eta^{16}(q^4)}{\eta^{24}(q^{2})},&&\beta:=16\frac{\eta^8(q^j)\eta^{16}(q^{4j})}{\eta^{24}(q^{2j})}.
\end{align*}
It is known that $\alpha$ and $\beta$ are algebraically related if $j$ is a positive integer, and we call such relations \textit{$j$th degree modular equations} (this is equivalent to Berndt's definition \cite[p.~212]{B93}).
When $j=3$ we appeal to \cite[p.~230]{B93}, to see that
\begin{equation*}
(\alpha\beta)^{1/4}+\left\{(1-\alpha)(1-\beta)\right\}^{1/4}=1.
\end{equation*}
Ramanujan parameterized the third-degree modular equation in terms of rational functions \cite[p.~230]{B93}:
\begin{align}\label{birational third degree modular equation}
\alpha=p\left(\frac{2+p}{1+2p}\right)^3,&&\beta=p^3\left(\frac{2+p}{1+2p}\right).
\end{align}
We easily solve \eqref{birational third degree modular equation} for $p$:
\begin{equation*}
p=2\frac{\eta^3\left(q^2\right)\eta^3\left(q^3\right)\eta^6\left(q^{12}\right)}{\eta\left(q\right)\eta^2\left(q^4\right)\eta^9\left(q^6\right)},
\end{equation*}
By the results in \cite[pg.~124-126]{B93}, we also have
\begin{align}
\left(\frac{\eta(q^2)}{\eta(q)}\right)^{24}=&\frac{\alpha}{16 (1-\alpha)^2},&\left(\frac{\eta(q^4)}{\eta(q^2)}\right)^{24}=&\frac{\alpha^2}{256 (1-\alpha)},\label{ratio of etas}\\
\left(\frac{\eta(q^2)}{\eta(-q)}\right)^{24}=&-\frac{\alpha(1-\alpha)}{16},&\left(\frac{\eta(q^4)}{\eta(-q^2)}\right)^{24}=&-\frac{\left(1-\sqrt{1-\alpha}\right)^6 \sqrt{1-\alpha}}{4 \alpha^4},\label{ratio of etas 2}
\end{align}
\begin{equation}\label{ratio of etas 3}
\left(\frac{\eta(q^8)}{\eta(q^4)}\right)^{24}=\frac{(1 - \sqrt{1 -\alpha})^6}{1024 \alpha^2 \sqrt{1 - \alpha }}.
\end{equation}
If $q\mapsto q^3$ in either \eqref{ratio of etas}, \eqref{ratio of etas 2}, or \eqref{ratio of etas 3}, then the identities remain valid as long as we replace $\alpha$ with $\beta$.
Applying \eqref{birational third degree modular equation}, \eqref{ratio of etas}, \eqref{ratio of etas 2}, and \eqref{ratio of etas 3} leads to:
\begin{align*}
t_1(q)=&\left(\frac{1-\alpha}{1-\beta}\right)^{1/2}\left(\frac{\beta}{\alpha}\right)^{1/4}+\left(\frac{1-\beta}{1-\alpha}\right)^{1/2}\left(\frac{\alpha}{\beta}\right)^{1/4},\\
=&\frac{2(1+p+p^2)(1+4p+p^2)}{(1-p^2)\sqrt{p(2+p)(1+2p)}}\\
s_2(-q)=&-\frac{16 (1-\alpha)^2}{\alpha},\\
s_4\left(q^2\right)=&\frac{16 (2-\alpha)^4}{ \alpha^2(1-\alpha)},\\
s_4\left(q^3\right)=&\frac{16 (1+\beta)^4}{\beta(1-\beta)^2},\\
s_4\left(-q^2\right)=&-\frac{4 \left(1-\sqrt{1-\alpha}\right)^2 \left(2-\alpha-6 \sqrt{1-\alpha}\right)^4}{\alpha^4\sqrt{1-\alpha}},\\
s_4\left(q^4\right)=&\frac{4 \left(1+\sqrt{1-\alpha}\right)^2 \left(2-\alpha+6 \sqrt{1-\alpha}\right)^4}{\alpha^4\sqrt{1-\alpha}}.
\end{align*}
After substituting the above parameterizations into \eqref{five term relation}, we can drop the dependence on $q$, and simply regard $\alpha$ and $\beta$ as functions of $p$.  It is possible to prove that $p=\frac{1}{2}\left(2+3 \sqrt{2}-2 \sqrt{3}-\sqrt{6}\right)$ when $q=e^{-\pi\sqrt{2/3}}$.  The derivation is an exercise in manipulating singular moduli \cite[p.~183]{B98}, and is easy to check numerically.
\begin{theorem}The following identity is valid:
\begin{equation}\label{g1h evaluated}
\begin{split}
g_1&\left(\frac{2(1+p+p^2)(1+4p+p^2)}{(1-p^2)\sqrt{p(2+p)(1+2p)}}\right)\\
&=-\frac{1}{4} f_2\left(-\frac{16 (1-\alpha)^2}{\alpha}\right)  + \frac{9}{20} f_4\left(\frac{16 (2-\alpha)^4}{ \alpha^2(1-\alpha)}\right)+\frac{3}{20} f_4\left(\frac{16 (1+\beta)^4}{\beta(1-\beta)^2}\right)\\
&\qquad + \frac{1}{10}f_4\left(-\frac{4 \left(1-\sqrt{1-\alpha}\right)^2 \left(2-\alpha-6 \sqrt{1-\alpha}\right)^4}{\alpha^4\sqrt{1-\alpha}}\right)\\
&\qquad- \frac{1}{5} f_4\left(\frac{4 \left(1+\sqrt{1-\alpha}\right)^2 \left(2-\alpha+6 \sqrt{1-\alpha}\right)^4}{\alpha^4\sqrt{1-\alpha}}\right),
\end{split}
\end{equation}
provided that
\begin{align*}
\alpha=p\left(\frac{2+p}{1+2p}\right)^3,&&\beta=p^3\left(\frac{2+p}{1+2p}\right),
\end{align*}
and $p\in[0,\frac{1}{2}\left(2+3 \sqrt{2}-2 \sqrt{3}-\sqrt{6}\right)]\approx[0,.16452\dots]$.
\end{theorem}
If we wish to compare formulas \eqref{numerical identity} and \eqref{g1h evaluated} numerically, it is sufficient to set $r=\frac{4p}{(1+p)^2}$ in \eqref{numerical identity}.
The left-hand side of \eqref{g1h evaluated} equals $g_1(6)$ when $p=\frac{1}{2}\left(2+3 \sqrt{2}-2 \sqrt{3}-\sqrt{6}\right)$.  Since $g_1(6)=\lambda_{sc}$, we obtain
\begin{equation}\label{lambdasc formula}
\begin{split}
\lambda_{sc}=&-\frac{1}{4} f_2\left(-\frac{16 (1-\alpha)^2}{\alpha}\right)  + \frac{9}{20} f_4\left(\frac{16 (2-\alpha)^4}{ \alpha^2(1-\alpha)}\right)+\frac{3}{20} f_4\left(\frac{16 (1+\beta)^4}{\beta(1-\beta)^2}\right)\\
&+ \frac{1}{10}f_4\left(-\frac{4 \left(1-\sqrt{1-\alpha}\right)^2 \left(2-\alpha-6 \sqrt{1-\alpha}\right)^4}{\alpha^4\sqrt{1-\alpha}}\right)\\
&- \frac{1}{5} f_4\left(\frac{4 \left(1+\sqrt{1-\alpha}\right)^2 \left(2-\alpha+6 \sqrt{1-\alpha}\right)^4}{\alpha^4\sqrt{1-\alpha}}\right)=1.673389302970196732283\ldots .
\end{split}
\end{equation}
We defined $f_{2}(u)$ and $f_{4}(u)$ in \eqref{fs} and \eqref{fs}, and $\alpha$ and $\beta$ equal
\begin{align*}
\alpha=&35-24 \sqrt{2}-20 \sqrt{3}+14 \sqrt{6},\\
\beta=&35+24 \sqrt{2}-20 \sqrt{3}-14 \sqrt{6}.
\end{align*}
Since all five of the hypergeometric functions are real-valued, we can drop the ``real part" notation from the hypergeometric functions in \eqref{fs} and \eqref{fs}. Note that a slight simplification arises if we observe that the argument of the third term above simplifies to 2304. No other terms obviously simplify.

\subsection{$L$-function formula  for the spanning tree constant.}
It is possible represent $\lambda_{sc}$ in terms of $L$-functions of modular forms \cite{RGIP}.  We have
\begin{equation*}
\lambda_{sc}=\frac{24\sqrt{6}}{\pi^3}L(f,3),
\end{equation*}
where $q=e^{2\pi i \tau}$, and
\begin{equation*}
f(\tau)=\frac{1}{2}\sum_{n,k=-\infty}^{\infty}\left(n^2-6k^2\right)q^{n^2+6k^2}-(2n^2-3k^2)q^{2n^2+3k^2}.
\end{equation*}
We can also represent $f(\tau)$ in terms of eta functions:
\begin{equation*}
\begin{split}
f(\tau)=&-\frac{ \eta(\tau)^2 \eta\left(4\tau\right)^7 \eta\left(6\tau\right)^3}{\eta\left(2\tau\right)^4 \eta\left(8\tau\right)^2}+\frac{\eta(\tau)^4 \eta\left(4\tau\right)^3 \eta\left(6\tau\right)^5 \eta\left(8\tau\right)^2}{\eta\left(2\tau\right)^4 \eta\left(3\tau\right)^2 \eta\left(12\tau\right)^2}\\
&+2\frac{\eta\left(2\tau\right)^7 \eta\left(8\tau\right)^2 \eta\left(12\tau\right)^3}{\eta(\tau)^2 \eta\left(4\tau\right)^4}+\frac{\eta(\tau)^2 \eta\left(2\tau\right)^3 \eta\left(8\tau\right)^4 \eta\left(12\tau\right)^5}{\eta\left(4\tau\right)^4 \eta\left(6\tau\right)^2 \eta\left(24\tau\right)^2}.
\end{split}
\end{equation*}
Bertin \cite{Bert2} has obtained some similar results along these lines.

\subsection{Diamond lattice}

For the diamond lattice, using the abbreviated notation $c_i \equiv \cos(k_i) \,\, i=1,2,3,$
 \begin{eqnarray*}
 T_{diam}(z) &=  &\log{4}+ \frac{1}{2\pi^3}\int_{0}^{\pi}  \int_{0}^{\pi} \int_{0}^{\pi}
 dk_1\, dk_2\, dk_3 \log\left (\frac{1}{z^2}-\frac{1}{4}(1+c_1c_2+c_1c_3+c_2c_3)\right ) \\ \nonumber
  & =&\log{4}-\frac{1}{2}\sum_{n \ge 1}\left ( \frac{z}{4}\right )^{2n}\frac{1}{n}\sum_{j=0}^n {n \choose j}^2{2j \choose j} {2n-2j \choose n-j}.
  \end{eqnarray*}

The argument of the logarithm above is non-negative for $z^2 \in [0,1],$ so can be replaced by the logarithm of the modulus for $z$ in that range. From eqn. (\ref{threem}) one then immediately recognizes the integral above as the Mahler measure $f(16/z^2)-4\log(2),$ so that $$T_{diam}(z)=f(16/z^2)-2\log(2).$$ Substitution into eqns (\ref{threehyper}) and (\ref{fs}) and straightforward algebraic manipulation then gives
\begin{eqnarray}\label{tdiam}
T_{diam}(z)&=&\frac{2}{5}\log(2)+\frac{4}{5}\log(4-z^2)-\frac{1}{10}\log(1-z^2) \\ \nonumber
&-&\frac{z^2}{40(1-z^2)^3}{_5F_4}\left(\substack{\frac{5}{3},\frac{3}{2},\frac{4}{3},1,1\\2,2,2,2}; \frac{-27z^4}{4(1-z^2)^3}\right ) \\ \nonumber
&-&\frac{4z^4}{5(4-z^2)^3}{_5F_4}\left(\substack{\frac{5}{3},\frac{3}{2},\frac{4}{3},1,1\\2,2,2,2}; \frac{27z^4}{(4-z^2)^3}\right ). \\ \nonumber
\end{eqnarray}

This expression is well defined for $|z| < 1.$ As $z \to 1,$ both a logarithmic term and the first hypergeometric term are singular. We wish to calculate
\begin{equation}\label{5F4 sum}
\begin{split}
\lim_{z\rightarrow 1}-\frac{1}{10}\log(1-z^2)-\frac{z^2}{40(1-z^2)^3}{_5F_4}\left(\substack{\frac{5}{3},\frac{3}{2},\frac{4}{3},1,1\\2,2,2,2};\frac{-27z^2}{4(1-z^2)^3}\right).
\end{split}
\end{equation}
If we express the $_5F_4$ function as an integral, then the limit becomes
\begin{equation*}
\begin{split}
=&\lim_{z\rightarrow 1}-\frac{1}{30}\log\frac{27z^2}{4}+\frac{1}{30}\int_{0}^{1}\frac{{_3F_2}\left(\substack{\frac{1}{3},\frac{1}{2},\frac{2}{3}\\1,1};-u\right)-1}{u}du+\frac{1}{30}\int_{1}^{\frac{27z^2}{4(1-z^2)^3}}\frac{{_3F_2}\left(\substack{\frac{1}{3},\frac{1}{2},\frac{2}{3}\\1,1};-u\right)}{u}du\\
=&-\frac{1}{30}\log\frac{27}{4}+\frac{1}{30}\int_{0}^{1}\frac{{_3F_2}\left(\substack{\frac{1}{3},\frac{1}{2},\frac{2}{3}\\1,1};-u\right)-1}{u}du+\frac{1}{30}\int_{1}^{\infty}\frac{{_3F_2}\left(\substack{\frac{1}{3},\frac{1}{2},\frac{2}{3}\\1,1};-u\right)}{u}du\\
=&\frac{2}{15}\log 2.
\end{split}
\end{equation*}
Mathematica can evaluate both definite integrals in terms of $_4F_3$ functions.  When the expressions are combined, all of the $_4F_3$ functions cancel out, and we are left with a multiple of $\log2$.

Thus we obtain for the diamond lattice spanning tree constant
$$\lambda_{diam} = \frac{8}{15}\log(2)+\frac{4}{5}\log(3)-\frac{4}{135}{_5F_4}\left(\substack{\frac{5}{3},\frac{3}{2},\frac{4}{3},1,1\\2,2,2,2}; 1\right ) = 1.2064599496517629\ldots $$

From eqn (2.25) in \cite{R09} we have the elegant result giving the spanning tree constant in terms of an $L$-function,
$$\lambda_{diam} = \frac{24}{\pi^2}L(\eta(2\tau)^3\eta(6\tau)^3,3)$$
where $\eta(\tau)$ is the Dedekind eta function.

For the 4d diamond lattice, since the LGF is identical to that of the 4d hyper-cubic lattice, and the two lattices have the same co-ordination numbers, it follows that the spanning tree constant of the 4d diamond lattice is the same as that of the 4d hypercubic lattice, given above.

 \subsection{Face-centred cubic lattice}

 After the completion of this work, we became aware of the work of G S Joyce \cite{J12} who studied both the STGF and LGF on the f.c.c. lattice making use of the relevant Mahler measure given by one of us in \cite{R09}. Accordingly, we have shortened this section and just given brief details of our derivation. A more extensive and thorough treatment of the f.c.c. case can be found in \cite{J12}.

The f.c.c. spanning tree constant was first given by Shrock and Wu \cite{SW00}, but there was an error in the structure function that they used. This was corrected by Chang and Shrock \cite{CS06} who correctly gave $\lambda_{fcc} \approx 2.41292.$ The spanning tree generating function is
  $$T_{fcc}(z) = \log{12} + \frac{1}{(2\pi)^3}\int_{-\pi}^{\pi} dk_1 \cdots  \int_{-\pi}^{\pi} dk_3 \log(1/z - \frac{1}{3}(\cos{k_1}\cos{k_2}+\cos{k_2}\cos{k_3}+\cos{k_3}\cos{k_1})).$$
From eqn ({\ref{threem}) we immediately identify $T_{fcc} $ as $f(4+12/z).$ Substitution into eqns (\ref{threehyper}) and (\ref{fs}) and similar algebraic manipulation as in the diamond case above then gives
\begin{eqnarray}\label{tfcc}
T_{fcc}(z)&=&2\log(2)+\frac{7}{5}\log(3)-\log(z) -\frac{2}{5}\log(3+z)-\frac{1}{5}\log(1-z) \\ \nonumber
&-&\frac{z(3+z)^2}{135(z-1)^3}{_5F_4}\left(\substack{\frac{5}{3},\frac{3}{2},\frac{4}{3},1,1\\2,2,2,2}; \frac{-z(3+z)^2}{(z-1)^3}\right ) \\ \nonumber
&-&\frac{2z^2(z+3)}{135}{_5F_4}\left(\substack{\frac{5}{3},\frac{3}{2},\frac{4}{3},1,1\\2,2,2,2}; \frac{z^2(z+3)}{4}\right ). \\ \nonumber
\end{eqnarray}
  As above, taking the limit as $z \to 1$ we obtain for the fcc lattice spanning tree constant
$$\lambda_{fcc} = 2\lambda_{diam}= \frac{16}{15}\log(2)+\frac{8}{5}\log(3)-\frac{8}{135}{_5F_4}\left(\substack{\frac{5}{3},\frac{3}{2},\frac{4}{3},1,1\\2,2,2,2}; 1\right ) = 2.4129198993035259\ldots $$
As for the diamond lattice case, from eqn (2.25) in \cite{R09} we have the result in terms of L-functions,
$$\lambda_{fcc} = \frac{48}{\pi^2}L\left(\eta(2\tau)^3\eta(6\tau)^3,3\right).$$

 Finally, for the four-dimensional fcc lattice we find by integration of the relevant LGF \cite{G09},
 $\lambda_{fcc} (4)\approx 3.14567\ldots .$
\section{Integral identities and hypergeometric identities}
Having shown the connection between STGFs and LGFs, we can immediately write down a number of integral identites which follow directly from the fact that the integral of the LGF is the STGF.  As two- and three-dimensional LGFs can be expressed as complete elliptic integrals of the first kind, and as the square of complete elliptic integrals of the first kind respectively, while STGFs are expressible in terms of hypergeometric functions, it follows that we can express the integrals of complete elliptic integrals and their square in terms of higher order hypergeometric functions, while differentiating the STGF result relates higher order hypergeometric functions to $_2F_1$ hypergeometric functions, or their square.

In this way a large number of identities can be produced. We have not yet determined how new and/or useful these results are, so we just give a couple of examples at this stage, thereby making the procedure clear. We will investigate the full family in greater detail subsequently. A simple, previously known, result is given above in equation (\ref{lamsq}). Less well-known, or possibly unknown results follow in the case of three-dimensional lattices. The simplest situation arises with the body-centred cubic case, by combining eqn (\ref{treeconst}) and the equation above for  $\lambda_c^{bcc}(3) $, giving the integral identity
$$ \int_0^1 \frac{\left [ \frac{2}{\pi}{\bf K}(k_2)\right ]^2-1}{z} dz
=  \frac{1}{16}  {_{5}F_{4}}\left (\substack {1,1,\frac{3}{2}, \cdots , \frac{3}{2}\\ 2,2, \cdots , 2};1\right ),
$$
 where $$k_2^2=\frac{1}{2} - \frac{1}{2}\sqrt{1-z^2}.$$
More generally,
$$ \int \frac{\left [ \frac{2}{\pi}{\bf K}(k_2)\right ]^2}{z} dz
=  \log{z}+\frac{z^2}{16}  {_{5}F_{4}}\left (\substack {1,1,\frac{3}{2}, \cdots , \frac{3}{2}\\ 2,2, \cdots , 2}; z^2\right ).
$$
Similar identities can be obtained for all the other two- and three-dimensional lattices.

For example, by combining eqn (\ref{treeconst}) and the equation above for  $\lambda_c^{sc}(3) $, we now have the  integral identity
\begin{equation}\label{lambdasc formula}
\begin{split}
\lambda_{sc}=&\int_0^1 \frac{\frac{1-9\xi^4}{(1-\xi)^3(1+3\xi)}\left [ \frac{2}{\pi}{\bf K}(k_1)\right ]^2-1}{z} dz
=  -\frac{1}{2}\sum_{l=1}^\infty \frac{1}{l}  {2l \choose l} {_3F_2}\left (\substack {\frac{1}{2},-l,-l\\1,1};4\right )\\
=&-\frac{1}{4} f_2\left(-\frac{16 (1-\alpha)^2}{\alpha}\right)  + \frac{9}{20} f_4\left(\frac{16 (2-\alpha)^4}{ \alpha^2(1-\alpha)}\right)+\frac{3}{20} f_4\left(\frac{16 (1+\beta)^4}{\beta(1-\beta)^2}\right)\\
&+ \frac{1}{10}f_4\left(-\frac{4 \left(1-\sqrt{1-\alpha}\right)^2 \left(2-\alpha-6 \sqrt{1-\alpha}\right)^4}{\alpha^4\sqrt{1-\alpha}}\right)\\
&- \frac{1}{5} f_4\left(\frac{4 \left(1+\sqrt{1-\alpha}\right)^2 \left(2-\alpha+6 \sqrt{1-\alpha}\right)^4}{\alpha^4\sqrt{1-\alpha}}\right),
\end{split}
\end{equation}
 where $$k_1^2=\frac{16\xi^3}{(1-\xi)^3(1+3\xi)}; $$  with $$ \xi=(1+\sqrt{1-z^2})^{-1/2}(1-\sqrt{1-z^2/9})^{1/2}$$
and several other identitites also follow immediately.

As another example, by combining eqn (\ref{treeconst}) and the equations above for  $\lambda_{fcc} $, we now have the (possibly new) result:
$$ \lambda_{fcc} = \int_0^1 \frac{\frac{(1+3\xi^2)^2}{(1-\xi)^3(1+3\xi)}\left [ \frac{2}{\pi}{\bf K}(k_3)\right ]^2-1}{z} dz
 = \frac{48}{\pi^2}L(\eta(2\tau)^3\eta(6\tau)^3,3)$$
 where $$k_3^2=\frac{16\xi^3}{(1-\xi)^3(1+3\xi)};  $$ and $$ \xi=(1+\sqrt{1-z})^{-1}(-1+\sqrt{1+3z}).$$

 An alternative expression for $\lambda_{fcc}$ is
 $$\lambda_{fcc}=\log {12}-\sum_{n>0}\frac{1}{n} \left ( \frac{1}{12} \right )^n \sum_{j=0}^n {n \choose j} (-4)^{n-j} \sum_{k=0}^j {j \choose k}^2 {2k \choose k} {2j-2k \choose j-k}.$$

 Note too that  $\lambda_{d} =  \lambda_{fcc}/2,$ giving rise to the identity
  \begin{eqnarray*}
 && \sum_{n>0}\frac{1}{n} \left ( \frac{1}{12} \right )^n \sum_{j=0}^n {n \choose j} (-4)^{n-j} \sum_{k=0}^j {j \choose k}^2 {2k \choose k} {2j-2k \choose j-k}-\log{\frac{3}{4}} \\ \nonumber
  &=  &\sum_{n \ge 1}\left ( \frac{1}{4}\right )^{2n}\frac{1}{n}\sum_{j=0}^n {n \choose j}^2{2j \choose j} {2n-2j \choose n-j}.
  \end{eqnarray*}

Going in the opposite direction, for the square lattice one can, for example, differentiate equation (\ref{treesq}), giving
\begin{eqnarray}
-z\frac{dT}{dz}&=&1+ \frac{z^2}{4}  {_{4}F_{3}}\left (\substack {1,1,\frac{3}{2},\frac{3}{2}\\ 2,2, 2}; z^2\right )+ \frac{9z^4}{128}  {_{4}F_{3}}\left (\substack {2,2,\frac{5}{2},\frac{5}{2}\\ 3,3,3};z^2\right )\\ \nonumber
&=&1+\frac{z^2}{4} + \frac{9z^4}{64}  {_{3}F_{2}}\left (\substack {1,\frac{5}{2},\frac{5}{2}\\ 3,3}; z^2\right ) = {_{2}F_{1}}\left (\substack {\frac{1}{2},\frac{1}{2}\\ 1}; z^2\right ).
\end{eqnarray}
The second equality follows by algebraic manipulation of the sum of the two higher  order hypergeometric functions, and the final equality follows as the r.h.s {\it is} the square lattice LGF.

A corresponding calculation based on the results above for the body-centred cubic lattice STGF and LGF gives
\begin{eqnarray}
-z\frac{dT}{dz}&=&1+ \frac{z^2}{8} +  \frac{27z^4}{512}{_{4}F_{3}}\left (\substack {1,\frac{5}{2},\frac{5}{2},\frac{5}{2}\\ 3,3,3}; z^2\right )
= {_{3}F_{2}}\left (\substack {\frac{1}{2},\frac{1}{2},\frac{1}{2}\\ 1,1}; z^2\right ).
\end{eqnarray}
Again, similar identities can be obtained for all the other two- and three-dimensional lattices.

\section{Connection between spanning trees and the Ising model and dimer coverings.}
There is a close connection between the spanning tree constant $\lambda$ and the free-energy of the Ising model at the critical temperature. Unfortunately, this seems to be true only for planar lattices (otherwise we would have some clue as to the structure of the free-energy of the 3-dimensional Ising model). The Onsager solution for the free-energy of the square-lattice Ising model is
\begin{equation}
F(v) = \log\left (\frac{2}{1-v^2}\right ) + \frac{1}{8\pi^2} \int_{-\pi}^{\pi} \int_{-\pi}^{\pi} \log[(1+v^2)^2-2v(1-v^2)(\cos x+\cos y)]dx.dy
\end{equation}
where $v = \tanh(\frac{J}{k_BT}).$ At the critical temperature, $v = v_c = \sqrt{2}-1,$ this simplifies to
\begin{equation}
F(v_c) = \frac{1}{2}\log(2) + \frac{1}{8\pi^2} \int_{-\pi}^{\pi} \int_{-\pi}^{\pi} \log[4-2(\cos{x}+\cos{y})]dx.dy = (\lambda_{sq} + \log{2})/2.
\end{equation}

If $b_n$ denotes the number of distinct dimer coverings of an $n \times n$ square lattice (with $n$ even), then, as shown by both Kasteleyn \cite{Kas} and by Fisher and Temperley \cite{FT},
$$b_n = 2^{n/2} \prod_{j=1}^{n/2} \prod_{k=1}^{n/2}\left ( \cos^2{\frac{j\pi}{n+1}} + \cos^2{\frac{k\pi}{n+1}}\right ).$$ In the infinite lattice limit,
$$\lim_{n \to \infty} \frac{1}{n} \log{b_n} = \frac{1}{16\pi^2} \int_{-\pi}^{\pi} \int_{-\pi}^{\pi} \log[4+2(\cos{x}+\cos{y})]dx.dy = \frac{G}{\pi},$$ where $G$ is Catalan's constant. This is very similar to the expression for $\lambda_{sq},$ apart from a (critical) sign change. However Temperley \cite{T74} pointed out that if one considers dimers on a $(2n-1) \times (2n-1)$ site lattice, with one boundary site removed, the relevant sign changes, and the integral agrees with that for the spanning tree constant. Later, Tzeng and Wu \cite{TW03} evaluated the dimer generating function for the one vacancy case, and showed that it was independent of the location of the vacant site. The Temperley bijection does not apply to cylindrical lattices however. For that situation, the relevant dimer generating function has been evaluated by Wu, Tzeng and Izmailian using Pfaffians \cite{WTI11}.

In \cite{S05}, Stienstra points out a connection between the partition function of certain dimer models and the $L$-functions of their spectral curves. This follows from his observation that the partition function per fundamental domain of a dimer model, as given by Kenyon, Okounkov and Sheffield \cite{KOS03} is in fact the logarithmic Mahler measure of the characteristic polynomial of that dimer model. Stienstra demonstrates the connection explicitly for three different dimer models.

\section{Conclusion}
We have derived the spanning tree constants for all the usual three-dimensional lattices in terms of $_5F_4$ hypergeometric functions. This has been made possible by establishing the connection between the integral representation of the spanning tree constant and the Mahler measure of an approporiate Laurent polynomial, which is closely related to the structure function of the underlying lattice.
We have introduced the notion of a spanning tree generating function, which gives the spanning tree constant as the value of the STGF at $z=1.$ We show the simple connection between the STGF and the lattice Green function, and so by comparing known results for the LGF with known and new results for the STGF, we are able to derive a number of integral and hypergeometric identities.
We have also expressed all of the spanning tree constants for both two- and three-dimensional lattices in terms of Dirichlet $L$-series.

\section*{Acknowledgements}
This project was conceptualised when I (AJG) read the seminal works of Fa Yueh Wu on spanning tree constants. Work started in earnest when I was a guest at the Mittag-Leffler Institute in the program {\em Discrete Probability} in 2009. I have subsequently benefited from discussions with several colleagues, including Larry Glasser, Christian Krattenthaler, Christoph Koutschan, Jan Stienstra, Ole Waarnar, Fred Wu, John Zucker and Wadim Zudilin. This work was supported by the Australian Research Council through grant DP1095291. I wish to express my gratitude to the above-named institutions and individuals.



\begin{thebibliography}{99}

\bibitem{BKW76}
R J Baxter, S B Kelland F Y Wu (1976)
J. Phys. A:Math. Gen. {\bf 9}, 397--406.

\bibitem{Bert}
M.\,J.~Bertin (2008)
\emph{J. Number Theory} {\bf 128}, 2890--2913.
\url{http://arxiv.org/abs/math/0501153}.

\bibitem{Bert2}
M.\,J.~Bertin (2008)
\url{http://arxiv.org/abs/0803.0413}.

\bibitem{Be89}
B.\,C.~Berndt,
\emph{Ramanujan's Notebooks, Part II}
(Springer-Verlag, New York, 1989).


\bibitem{B93}
N L Biggs (1993)
{\it Algebraic Graph Theory}, (2nd edition). (1993)
Cambridge University Press, Cambridge.

\bibitem{B81} D W Boyd (1981)
Canad. Math. Bull. {\bf 24}, 453-469.

\bibitem{B98} D W Boyd (1998)
Experiment. Math. {\bf 7}, 37--82.


\bibitem{BP93}
R M Burton and R Pemantle (1993)
Ann. Probab. {\bf 21}, 1329--1371.
\bibitem{CTYZ}
H H Chan, Y Tanigawa, Y Yang and W. Zudilin (2011)
Adv. in Math. {\bf 228}, 1294--1314

\bibitem{C08}
S-C Chang (2009)
J. Phys. A:Math. Gen. {\bf 42}, 015208.

\bibitem{CW06}
S-C Chang and W Wang (2006)
J. Phys. A:Math. Gen. {\bf 39}, 10263--75.

\bibitem{CS06}
S-C Chang and R Shrock (2006)
J. Phys. A:Math. Gen. {\bf 39}, 5653--5658.

\bibitem{FL03}
J L Felker and R Lyons (2003)
J. Phys. A:Math. Gen. {\bf 36}, 8361--8365.

\bibitem{FK}
C M Fortuin and P W Kasteleyn (1972),
Physica {\bf57}, 536--64.


\bibitem{GL05}
M L Glasser and G Lamb (2005)
J. Phys. A:Math. Gen. {\bf 38}, L471--5.

\bibitem{GW05}
M L Glasser and F Y Wu (2005)
Ramanujan J {\bf 10}, 205-14.

\bibitem{G09}
A J Guttmann (2009)
J. Phys. A:Math. Gen. {\bf 42}, 232001 (6pp).

\bibitem{G10}
A J Guttmann (2010)
J. Phys. A:Math. Gen. {\bf 43}, 305205 (26pp).



\bibitem{H95}
B D Hughes (1995)
{\it Random Walks and Random Environments, Vol. 1: Random Walks}
Clarendon Press: Oxford.

\bibitem{J73}
G S Joyce (1973)
Phil. Trans. Royal Soc. London, {\bf A273}, 583--610.

\bibitem{J98}
G S Joyce (1998)
J. Phys. A:Math. Gen. {\bf 31}, 5105--15.

\bibitem{J01}
G S Joyce (2001)
J. Phys. A:Math. Gen. {\bf 34}, 3831--39.

\bibitem{J12}
G S Joyce (2012)
J. Phys. A:Math. Gen. {\bf 45}, 285001 (12pp).

\bibitem{JDZ98}
G S Joyce, R T Delves and I J Zucker (1998)
J. Phys. A:Math. Gen. {\bf 31}, 1781--1790.

\bibitem{JZ01}
G S Joyce and I J Zucker, (2001)
J. Phys. A:Math. Gen. {\bf 34}, 7349--7354.

\bibitem{Kas} P Kasteleyn (1961)
Physics {\bf 27}, 1209--1225.

\bibitem{KOS03}
R A Kenyon, A Okounkov and S Sheffield (2006)
Ann. Math. {\bf 163},  no. 3, 1019--1056,
arXiv:math-ph/0311005

\bibitem{LR} M Lalin and M D Rogers (2007)
Algebra Number Theory, {\bf 1}, no. 1, 87--117.

\bibitem{L05}
R Lyons (2005)
Combin. Probab. and Comput. {\bf 14}, 491--522.

\bibitem{MMW60}
A A Maradudin, E W Montroll and G H Weiss R Herman and H W Milnes (1960)
M\'emoires Acad. Roy. de Belg. (Classe des Sciences), Tome XIV, Fasicule 7 No 1709.

\bibitem{RV} F. Rodriguez-Villegas (1999)
 Modular Mahler measures I,
Topics in number theory (University Park, PA, 1997), 17--48, Math.
Appl., 467, Kluwer Acad. Publ., Dordrecht.

\bibitem{R09} M D Rogers (2009)
Ramanujan J {\bf18}, 327--340.

\bibitem{R12} M Rogers (2011)
IMRN, no. 17, 4027--4058.

\bibitem{RGIP} M Rogers (2012), In progress.

\bibitem{R87}
A Rosengren (1987)
J. Phys. A:Math. Gen. {\bf 20}, L923--927.

\bibitem{S12}
D Samart (2012)
arXiv 1205.4803v1.

\bibitem{SW00}
R Shrock and F Y Wu (2000)
J. Phys. A:Math. Gen. {\bf 33}, 3881--3902.

\bibitem{St76}
M Stephen (1976)
Phys. Letts {\bf 56 A}, 149--50.

\bibitem{S05}
J Stienstra (2006)
Proceedings of {\it Workshop on Calabi-Yau Varieties and Mirror Symmetry}
Edited by N. Yui, S.-T. Yau, J.D. Lewis. Providence, Am. Math. Soc. 576p. (AMS/IP studies in advanced mathematics 38)
arXiv:math/0502197v1

\bibitem{T74} H N V Temperley (1974)
{\it Combinatorics: Proceedings of the British Combinatorial Conferences, 1973}, London Mathematical Society Lecture Note Series \# 13, 202--204.

\bibitem{FT}  H N V Temperley and M E Fisher (1961),
Phil. Mag. {\bf 6}, 1061--63.

\bibitem{TW00}
W-J Tzeng and F Y Wu (2000),
Appl. Math. Letts. {\bf 13}, 19--25.

\bibitem{TW03}
W-J Tzeng and F Y Wu (2003),
J. Stat. Phys. {\bf 110}, 671--89.

\bibitem{We} E W Weisstein,
\url{http://mathworld.wolfram.com/DirichletL-Series.html}

\bibitem{W77}
F Y Wu (1977)
J. Phys. A:Math. Gen. {\bf 10}, L113--115.

\bibitem{WTI11}
F Y Wu, W-J Tzeng and N Sh Izmailian (2011),
Phys. Rev E {\bf 83}, 011106 (6pp).

\bibitem{Z11} I. J. Zucker (2011)
 J. Stat. Phys, {\bf 145}, 591--612.















\end{thebibliography}
\end{document}